\documentclass[twocolumn]{article}
\usepackage{arxiv}  %

\usepackage{cite}
\usepackage{amsmath,amssymb,amsfonts}
\usepackage{algorithmic}
\usepackage{graphicx}
\usepackage{textcomp}
\usepackage{xcolor}
\usepackage{fancyhdr}
\usepackage{lipsum} 
\usepackage{booktabs}
\usepackage{array}
\usepackage{multirow}
\usepackage[T1]{fontenc}
\usepackage[inline]{enumitem}

\usepackage{tikz}
\usetikzlibrary{arrows.meta}
\usetikzlibrary{decorations.pathreplacing}
\usetikzlibrary{decorations}
\usepackage{pgfplots}
\pgfplotsset{compat=1.9}
\usepgfplotslibrary{groupplots}

\newcolumntype{M}[1]{>{\centering\arraybackslash}m{#1}}

\usepackage[pdfencoding=auto, pdfusetitle]{hyperref}
\hypersetup{
  hidelinks,
  urlcolor=red,
  pdftitle={Emotion Recognition with Pre-Trained Transformers Using Multimodal Signals},
  pdfauthor={Juan Vazquez-Rodriguez, Grégoire Lefebvre, Julien Cumin, James L. Crowley},
  pdfkeywords={Affective Computing, Multimodal Emotion Recognition, Machine Learning},
}

\title{Emotion Recognition with Pre-Trained Transformers Using Multimodal Signals}

\author{Juan Vazquez-Rodriguez$^{1,2}$, Grégoire Lefebvre$^{1}$, Julien Cumin$^{1}$, James L. Crowley$^2$\vspace{0.1cm}\\
{$^1$ Orange Labs, Grenoble, France}\\
{$^2$ Univ. Grenoble Alpes, CNRS, Grenoble INP, LIG, 38000 Grenoble, France}\\ %
}

\begin{document}

\twocolumn[{%
  \begin{@twocolumnfalse}
    \maketitle
  \end{@twocolumnfalse}
}]

\setcounter{footnote}{0}

\begin{abstract}
In this paper, we address the problem of multimodal emotion recognition from multiple physiological signals. We demonstrate that a Transformer-based approach is suitable for this task. In addition, we present how such models may be pre-trained in a multimodal scenario to improve emotion recognition performances. We evaluate the benefits of using multimodal inputs and pre-training with our approach on a state-of-the-art dataset.

\end{abstract}
\keywords{Affective Computing, Multimodal Emotion Recognition, Machine Learning.}

\section{Introduction} \label{section:intro}

The increasing availability of mass-market wearable devices equipped with sensors for physiological signals provides new possibilities for monitoring the emotional health and well-being of users \cite{abdatHumanComputerInteractionUsing2011}. Although less reliable than medical-grade sensors, signals from wearable sensors like electrocardiograms (ECG) and electroencephalograms (EEG) can be combined to provide estimates of the emotional state of users.

In this work, we report on experiments with a Transformer-based approach for interpreting emotional state from different physiological signals obtained from wearable devices. We explore the estimation of emotional state from individual sensor modalities, including EEG and ECG, and show that 
fusing the two modalities leads to better results indicating that these modalities convey complementary information.  Furthermore, we demonstrate that a Transformer-based approach can be used to provide reliable estimates of emotional state from such signals. 

We center our work on recognizing emotions from ECG and EEG signals. There are other works that address multimodal emotion recognition \cite{rahmanIntegratingMultimodalInformation2020, siriwardhanaMultimodalEmotionRecognition2020, siriwardhanaJointlyFineTuningBERTlike2020, khareSelfSupervisedLearningCrossModal2021}, but the majority use signals such as images, sound and text, and not physiological signals. Although some authors have explored the use of physiological signals for emotion recognition \cite{correaAMIGOSDatasetAffect2018, siddharthUtilizingDeepLearning2019, rossUnsupervisedMultimodalRepresentation2021}, such signals have received less attention than other sensing modalities.

A common problem when addressing the task of emotion recognition is the lack of labeled data to effectively train deep-learning models \cite{correaAMIGOSDatasetAffect2018}. A possible approach to address this problem is the use of unsupervised pre-training techniques \cite{erhanWhyDoesUnsupervised2010}. However, pre-training with multiple signal modalities raises additional challenges. In this work, we investigate the use of a late-fusion approach, where we pre-train and fine-tune different single-modality models,
and then combine the outputs of the individual models
to obtain a fused feature that can be used to perform emotion prediction.

We use a Transformer \cite{vaswaniAttentionAllYou2017} to process physiological signals. The Transformer was originally developed for Natural Language Processing (NLP) tasks, with the intent of processing sequences of words.
Given that physiological signals are sequences of values,
the Transformer can be adapted for physiological signal processing
\cite{yanFusingTransformerModel2019}.
Transformers employ a learned attention mechanism to dynamically score the relevance of different parts of an input according to context. Attention-based processing is appropriate for processing physiological signals, as some parts of a signal may convey more information than other parts depending on the task and context.  Another advantage of using a Transformer is that we can benefit from a very successful pre-training technique described in BERT \cite{devlinBERTPretrainingDeep2019} and developed for NLP tasks, which we can adapt to our needs.
This pre-training strategy has been successfully adapted to other domains like Computer Vision \cite{sunVideoBERTJointModel2019}, Speech Processing \cite{huangSpeechRecognitionSimply2021} and Affective Computing \cite{vazquez-rodriguezTransformerBasedSelfSupervisedLearning2022}.

The main contributions of this paper are:
\begin{enumerate}
    \item We present a technique for recognizing emotions from multimodal physiological signals using a Transformer.
    \item We describe a method to pre-train the Transformer for recognizing emotions from multimodal physiological signals.
    \item We provide results from experiments that show that a multimodal pre-training strategy is effective for improving emotion recognition performances.
\end{enumerate}

\section{Related Work}

Contrary to traditional techniques like Gaussian naive Bayes \cite{gjoreskiInterdomainStudyArousal2018}, k-Nearest Neighbours \cite{shuWearableEmotionRecognition2020} and Support Vector Machines \cite{ correaAMIGOSDatasetAffect2018}, deep-learning may be used to recognize emotions directly from sensor signals without a need to design feature descriptors.  This is particularly useful for the recognition of emotions from physiological signals where there are no well-established feature descriptors for signal encoding.  

An example of a deep-learning approach is provided by the work of Santamaria et al. \cite{santamaria-granadosUsingDeepConvolutional2019}, where they employ models based on Convolutional Neural Networks (CNN) to perform emotion recognition. Another example is the work of Harper and Southern \cite{harperBayesianDeepLearning2020}, who use a combination of Recurrent Neural Networks (RNN) and CNNs. The Transformer\cite{vaswaniAttentionAllYou2017}, which uses stacked layers of self-attention,  has recently emerged as a powerful alternative to Convolutional and Recurrent Networks. 
In this work, we are interested in whether a Transformer architecture can be an effective tool to recognize emotions from multiple physiological signals. 

A variety of authors have explored deep-learning models for emotion recognition using multimodal signals.  Most of these works use images, audio and/or text as inputs \cite{rahmanIntegratingMultimodalInformation2020, siriwardhanaMultimodalEmotionRecognition2020, siriwardhanaJointlyFineTuningBERTlike2020, khareSelfSupervisedLearningCrossModal2021, chenMultimodalConditionalAttention2016, ghalebMultimodalTemporalPerception2019}. In a few cases, physiological signals have been used to improve recognition from image, audio and text \cite{choiMultimodalAttentionNetwork2020, xingExploitingEEGSignals2019, matsudaEmoTourMultimodalEmotion2018}. A few authors have described the use of multiple physiological signal modalities \cite{correaAMIGOSDatasetAffect2018, siddharthUtilizingDeepLearning2019, rossUnsupervisedMultimodalRepresentation2021}. 
These works consistently demonstrate the benefits of exploiting multiple signal modalities to improve the performance of emotion recognition.

Some multimodal approaches employ pre-training techniques to improve their results. The authors of \cite{rahmanIntegratingMultimodalInformation2020, siriwardhanaMultimodalEmotionRecognition2020, siriwardhanaJointlyFineTuningBERTlike2020, khareSelfSupervisedLearningCrossModal2021} develop models based on Transformers, using images, audio and text to recognize emotions. Rahman et al. \cite{rahmanIntegratingMultimodalInformation2020} report on the use of BERT \cite{devlinBERTPretrainingDeep2019}, a Transformer-based model pre-trained for NLP tasks, to process text, incorporating visual and audio modalities in a middle-fusion process. Siriwardhana et al. \cite{siriwardhanaMultimodalEmotionRecognition2020, siriwardhanaJointlyFineTuningBERTlike2020} describes the use of pre-trained models to extract features from visual, audio and text modalities, and followed by a cross-modal Transformer \cite{tsaiMultimodalTransformerUnaligned2019} to combine these different features. Khare et al. \cite{khareSelfSupervisedLearningCrossModal2021} use a BERT-like approach, masking some words in the input text, along with the audio and visual parts that correspond to those words, and then pre-training the model by predicting the masked words.

Some authors have explored pre-training approaches to recognize emotions from physiological signals. Sarkar and Etemad \cite{sarkarSelfSupervisedLearningECGBased2020} pre-train their CNN-based model by modifying the input signal with different transformations, such as adding noise or upscaling, and then pre-training the model to predict which transformation was used on the input signal. Vazquez-Rodriguez et al. \cite{vazquez-rodriguezTransformerBasedSelfSupervisedLearning2022} use a model based on a Transformer, which is pre-trained by masking some values in the input signal and trying to predict those masked values. These approaches only employ a single physiological modality.

Other works have explored pre-training for multimodal emotion recognition from physiological signals. Ross et al. \cite{rossUnsupervisedMultimodalRepresentation2021} and Liu et al. \cite{liuEmotionRecognitionUsing2016} use Variational Autoencoders (VAE) for each modality to extract representations from each physiological signal. The representations of all signals are then concatenated, and a second-level classifier is trained to predict emotion. Yang and Lee \cite{yangAttributeinvariantVariationalLearning2019} also use a VAE, but they use a single VAE by concatenating the different signals at the input level. However, unlike the use of the Transformer, these approaches do not benefit from self-attention. 

We can see that there is a void in the research regarding different approaches for emotion recognition: multimodal pre-training approaches are not typically used on physiological signals; conversely, pre-training approaches for physiological signals are usually single-modality; finally, the few multimodal pre-trained approaches for physiological signals we surveyed don’t use attention-based models. In this work, we thus propose to investigate the use of attention-based models like the Transformer to recognize emotions in a multimodal physiological scenario, with the use of pre-training techniques.

\section{Approach} \label{section:apporach}

\subsection{Multimodal Emotion Recognition}
When performing a classification task using multiple modalities, an important problem is to determine the processing level at which different signal modalities should be combined or \textit{fused}. One option is \textit{early-fusion}, where the combination of input features is used as input for the model. A second option is to do \textit{late-fusion}, where the outputs of single-modality models are combined and a second-level model is trained to perform the classification. A third option is a compromise between these two extremes: \textit{middle-fusion} that combines features from intermediate layers of the models.

Early-fusion can be provided by simply concatenating the input signals in the temporal dimension, to form a single (longer) sequence. The problem with this approach is that the computational complexity of the Transformer is $O(n^2)$, where $n$ is the length of the input. Therefore, a late-fusion approach has the advantage of simplifying the training process: several single-modality models can be trained one by one on less powerful hardware than the one required to train a single, more computationally expensive multimodal model. Then, if the single-modality models are frozen, the second-level model of the late-fusion approach can also be easily trained without using very powerful hardware resources. A similar reasoning can be applied to see the advantages of late-fusion over middle-fusion in this scenario.

Another major difficulty with early-fusion is that it hinders the benefits of pre-training techniques. With pre-training, we seek to use many different datasets, not necessarily related to the task of emotion recognition, to obtain a more general representation of the information from the different modalities. However,  early-fusion limits pre-training to datasets that include all of the targeted modalities, thus severely limiting the availability of datasets that can be used. With late-fusion, one can pre-train each single-modality model independently from one another, with potentially different datasets.

In our case, we are interested in a multimodal approach that allows us to use pre-training techniques to improve the performance of the model. In particular, we are interested in attention-based models such as the Transformer \cite{vaswaniAttentionAllYou2017}, and in pre-training techniques similar to the ones used in BERT \cite{devlinBERTPretrainingDeep2019}. 

In this work, we explore the use of a late fusion approach. This allows self-supervised pre-training for individual sensor modalities by reconstructing masked values in the input signal, similar to what is done in BERT \cite{devlinBERTPretrainingDeep2019}.
We refer to this as Masked Value Prediction or simply MVP.

With this approach, recognition training is performed in two steps. In the first step, we train two single-modality models: one to recognize emotions from electrocardiogram (ECG) signals, and one to recognize emotions from electroencephalogram (EEG) signals. Both models are trained separately, using MVP pre-training. In the second step, we concatenate the outputs of the single-modality models and use this combined representation to train a Fully-Connected Network (FCN) to recognize emotions.
While other fusion approaches may be possible like max or average pooling, or majority voting, concatenation has been found to provide a simple and effective technique for use with Transformers. 

In the rest of this section, we provide details about the ECG and EEG single-modality models, and the fused model that combines both signals.

\subsection{ECG Single-Modality Emotion Recognition} \label{section:single-modality}

\begin{figure*}[tb]
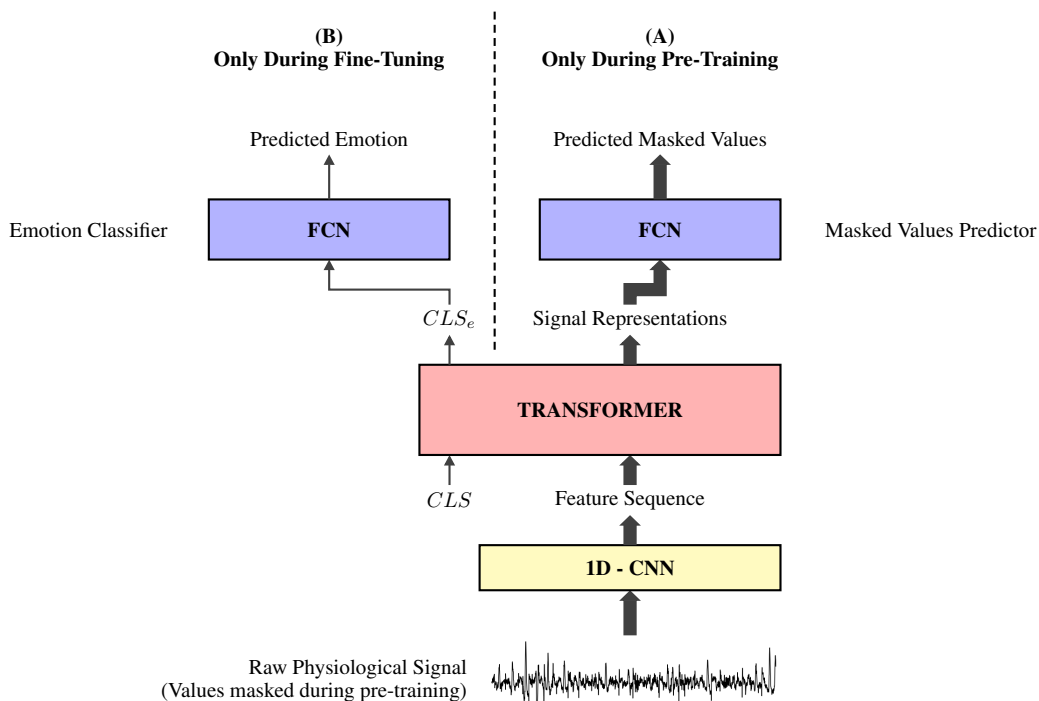

\centering
\include{single_modality}
\vspace*{-7.5mm}
\caption{Single-Modality Emotion Recognizer: The raw signal is encoded by a 1D-CNN and processed with a Transformer. First, the model is pre-trained by masking some values of the unlabeled input signal and then predicting those masked values (Part A). Then, labeled data is used to fine-tune the model in a supervised way (Part B).}
\label{fig:single_modality}
\end{figure*}

For our ECG emotion recognizer, we employ the approach described in \cite{vazquez-rodriguezTransformerBasedSelfSupervisedLearning2022} and depicted in Figure \ref{fig:single_modality}. In this approach, a pre-training step is first used prior to fine-tuning the model to improve its performance. We provide a brief description of this approach in the remainder of this subsection. We refer the reader to the original paper for further details.

The ECG single-modality emotion recognition model from \cite{vazquez-rodriguezTransformerBasedSelfSupervisedLearning2022} is based on the Transformer \cite{vaswaniAttentionAllYou2017}. The Transformer is an architecture capable of incorporating contextualized information thanks to its self-attention mechanisms. As shown in Figure \ref{fig:single_modality}, the model first encodes ECG signals using a 1D Convolutional Neural Network (1D-CNN) to obtain a sequence of features that represent the input signal. Then, similar to BERT \cite{devlinBERTPretrainingDeep2019}, a classification token named $CLS$ is added to the beginning of the sequence of features. In the next step, the feature sequence appended with the $CLS$ token is fed into a Transformer, which produces contextualized representations of the signal. Then, the process follows one of two paths, depending on if we are in the pre-training phase or the fine-tuning phase.

\subsubsection{Pre-Training Phase} \label{section:pretrain}
The pre-training task consists in using MVP, that is, masking some points from the original signal, and then predicting those masked points with the help of a Fully-Connected Network (FCN) placed on top of the Transformer, as can be seen in path A of Figure \ref{fig:single_modality}. Since this task is self-supervised, we do not need labeled data for this phase. 

\subsubsection{Fine-Tuning Phase}
After the model has been pre-trained, it is fine-tuned to recognize emotions. As depicted in path B of Figure \ref{fig:single_modality}, the vector $CLS_e$, which is the representation of the $CLS$ token, is used as input of an FCN that functions as a classifier to predict the emotion. Starting with the same pre-trained weights, the model is fine-tuned twice: one to predict arousal and another to predict valence. Thus, \textit{Predicted Emotion} in Figure \ref{fig:single_modality} refers to either arousal or valence. During this phase, all the parameters of the model, including the Transformer and the 1D-CNN parameters, are fine-tuned. This phase is supervised, therefore labeled data is employed.

\subsection{EEG Single-Modality Emotion Recognition}
\label{section:eeg_model}
To build our EEG single-modality emotion recognition model, we adapt the ECG model from \cite{vazquez-rodriguezTransformerBasedSelfSupervisedLearning2022} and described in the previous subsection, to accept EEG signals.

First, we need to take into account that the ECG signal has only one channel, while the EEG signal is typically multi-channel. Therefore, the 1D-CNN that encodes the raw signal has to be changed from having one-channel input to having as many inputs as the number of channels present in the EEG signal, while the rest of the structure of the 1D-CNN remains the same. The shape of the output from the 1D-CNN encoder remains similar to the original ECG model, and thus we can keep the same Transformer architecture.

Secondly, during the pre-training of the model, we mask the same temporal segments of the EEG signal across all the different channels. The size of the output layer of the FCN that works as masked-point predictor has to be the same as the number of input channels. This way, each predicted output value corresponds to the (masked) value of each channel.

Aside from these changes, no other adaptation is needed for fine-tuning the EEG model, as we employ the $CLS_e$ vector as input for the FCN used as a classifier to predict the emotion, as it is done in the ECG model.

\subsection{Fused-Signals Emotion Recognition Model} \label{section:fusion_estrategy}

\begin{figure}[tb]
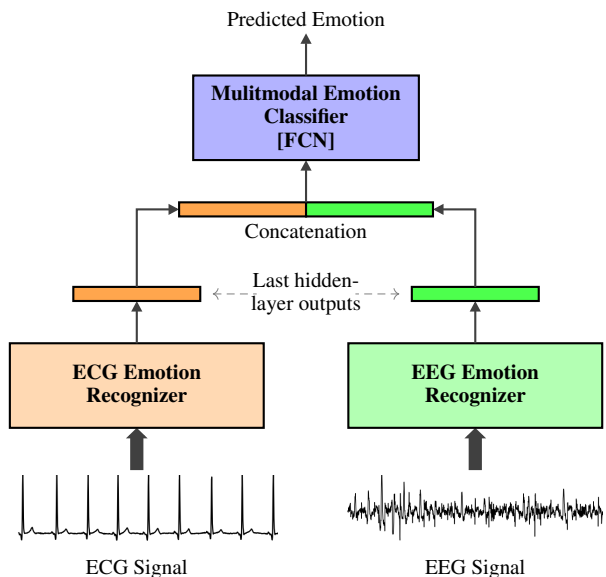

\centering
\include{fused_model}
\vspace*{-7.5mm}
\caption{Fused Model. Late-fusion is used to combine the ECG and EEG signals. The outputs of the last layer from both single-modality models are concatenated, and then used as input to an FCN that performs the emotion prediction.}
\label{fig:fused_model}
\end{figure}

We use late-fusion to fuse the ECG and EEG signals, using the outputs from each of the trained single-modality emotion recognizers. As depicted in Figure \ref{fig:fused_model}, we take the output of the last hidden layer of the FCN (not from the output layer) of each single-modality model, and we concatenate those outputs to form our combined features. Although other methods might be considered to do this fusion, for example using average pooling of the outputs of the single-modality recognizers, we employ concatenation because it allows us to have single-modality models with outputs of different sizes. This is convenient because this way we can choose without constraints the output sizes that make the fused model perform the best.

To perform the emotion prediction from the fused modalities, we use another FCN that we refer to as \textit{Multimodal Emotion Classifier} in Figure \ref{fig:fused_model}. For this step, we freeze the weights of the single-modality emotion recognizers, and only train the top FCN. More precisely, we train two FCNs: one to predict arousal and another to predict valence. This means that in Figure \ref{fig:fused_model} \textit{Predicted Emotion} refers to either arousal or valence.

\section{Experiments} \label{section:experiments}

We evaluate our model on the task of binary emotion prediction, that is to predict high and low levels of arousal and valence, from multimodal physiological signals. 

In this section, we describe our experimental setup, giving details about the datasets and the hyper-parameters for both EEG and fused-based models.

\subsection{Datasets}
\label{section:datasets}
To train and evaluate our models, we used the AMIGOS dataset \cite{correaAMIGOSDatasetAffect2018}. This dataset includes data from 40 subjects, where emotions were induced by making the subjects watch emotional videos. A total of 37 subjects watched 20 videos, while 3 subjects watched only 16. After watching each video, the subjects filled out a self-assessment form where they rated on a scale of 1 to 9 their levels of arousal and valence. We use the results of this self-assessment as labels in our experiments. Since we are interested in binary emotion classification, we use the average value as a threshold to obtain high and low classes of arousal and valence.  In total, there are around 65 hours of data in the AMIGOS dataset. 

The AMIGOS dataset includes both ECG and EEG signals. We treat ECG as a single-channel signal, and we use the signal taken from the left arm. For EEG, we use 10 channels: F7, F3, T7, P7, O1, O2, P8, T8, F4, F8. We chose those 10 channels from the 14 available because these channels are also present in the datasets that we use for pre-training, which we will describe shortly. We use the provided 128Hz down-sampled signals. ECG signals are filtered with a low-pass filter with a cut-off frequency of 60Hz, and EEG signals are filtered with a band-pass filter with frequencies between 4.0 and 45.0Hz. 

\subsubsection{Data for pre-training}
As described in Section \ref{section:single-modality}, the first step in training a single-modality emotion recognizer is pre-training. To pre-train the EEG emotion recognizer, we gathered EEG data that does not necessarily include labels of emotion. We use the following datasets: WAY-EEG-GAL \cite{luciwMultichannelEEGRecordings2014}, BCI2000 \cite{schalkBCI2000GeneralpurposeBraincomputer2004, goldbergerPhysioBankPhysioToolkitPhysioNet2000}, and Large-EEG-BCI \cite{kayaLargeElectroencephalographicMotor2018}. These datasets were gathered to develop Brain-Computer Interfaces, so they do not include any labels related to emotions. We also use parts of AMIGOS in the pre-training step, taking care of not using the same samples to pre-train and evaluate the model, for each fold of cross-validation. The quantity of data that we gathered to pre-train the EEG model is comparable to the data used to pre-train the ECG model in \cite{vazquez-rodriguezTransformerBasedSelfSupervisedLearning2022}: in total there are around 195 hours of EEG data available for pre-training, while the ECG model was pre-trained with around 230 hours of data.

\subsubsection{Signal pre-processing} Much like \cite{vazquez-rodriguezTransformerBasedSelfSupervisedLearning2022} for ECG data, we filter EEG signals using an 8\textsuperscript{th} order Butterworth band-pass filter, with cut-off frequencies of 0.8Hz and 50Hz. We also downsample signals to 128Hz. In addition, we normalize signals with zero-mean and unit-variance for each subject. Finally, we split each signal into 10-second segments. Each segment is used as a sample in our experiments, as in other state-of-the-art works \cite{correaAMIGOSDatasetAffect2018, sarkarSelfSupervisedLearningECGBased2020, rossUnsupervisedMultimodalRepresentation2021}, \cite{vazquez-rodriguezTransformerBasedSelfSupervisedLearning2022}.

\subsection{ECG Emotion Recognizer}

The ECG emotion recognition model follows the architecture presented by Vazquez-Rodriguez et al. in \cite{vazquez-rodriguezTransformerBasedSelfSupervisedLearning2022}. This model was fine-tuned and evaluated on AMIGOS as presented in Section \ref{section:datasets}. This model was parameterized and pre-trained as described in \cite{vazquez-rodriguezTransformerBasedSelfSupervisedLearning2022}.

\subsection{EEG Emotion Recognizer}
The EEG emotion recognition model follows the architecture described in Section \ref{section:eeg_model}. The 1D-CNN is composed of three layers with kernel sizes (65, 33, 17), with the number of channels equal to (64, 128, 256), with stride 1 in all the layers, and using the ReLU activation function. With this configuration, the size of the receptive field is 113 input points, which is equivalent to 0.88s. Based on preliminary studies, we believe that this receptive field size is suitable for EEG signals. For the Transformer, the number of layers is 2 and the number of heads is also 2, with a hidden size of 256. 

For pre-training, the FCN used to predict masked points has one hidden layer with size 128 and ReLU activation, and an output layer that gives 10 outputs values, where each output value corresponds to the predicted value of each masked EEG channel. The mean squared error between the real and the predicted values is used as loss. We pre-train the model for 500 epochs, warming up the learning rate during the first 30 epochs up to 0.0005, and then use linear decay. We use Adam optimization with $\beta_1 = 0.9$, $\beta_2 = 0.999$ and $L_2$ weight decay of 0.005. A dropout value of 0.1 is used in the Transformer.

For fine-tuning, the FCN that predicts binary emotions has one hidden layer with a size of 64 and ReLU activation functions. An output layer is used to project the output to a single value, that corresponds to the prediction of the emotion. Two different networks are fine-tuned, one to predict arousal and another to predict valence. The models are fine-tuned using binary cross-entropy loss for 100 epochs, starting with a learning rate of 0.0001 and decreasing by a factor of 0.65 every 45 epochs. Adam optimization is used, with $\beta_1 = 0.9$, $\beta_2 = 0.999$ and $L_2$ weight decay of 0.00001. A dropout value of 0.6 is used in the FCN that predicts emotions.

These hyper-parameters were optimized using the Ray Tune toolkit \cite{liawTuneResearchPlatform2018} on validation data extracted from AMIGOS, for each fold of cross-validation.

\subsection{Multimodal Emotion Recognizer}

The multimodal emotion recognition model follows the architecture described in Section \ref{section:fusion_estrategy}. The FCN has two hidden layers of size 64 and 32, and an output layer that projects the result to a single value used to predict the binary emotion. As we did for the single-modality models, we train one model to predict arousal and another to predict valence. The activation function used is ReLU. This network is trained for 52 epochs, starting with a learning rate of 0.00001 and decaying it every 20 epochs with a factor of 0.65. A dropout value of 0.1 is used during the training of this network. We employ Adam optimization with $\beta_1 = 0.9$, $\beta_2 = 0.999$ and $L_2$ weight decay of 0.00001. We also use the Ray Tune toolkit to optimize these hyper-parameters, as we did for the EEG emotion recognizer.

\section{Results} \label{section:results}

In this section, we discuss the results of the different experiments we performed to evaluate our model for emotion recognition on the AMIGOS dataset. We use the mean accuracy and F1 score between the two predicted classes as metrics, averaged across 10 folds of cross-validation. We also report a two-sided 95\% confidence interval, calculated with a t-distribution with 9 degrees of freedom.

\subsection{EEG single-modality Emotion Recognition}

\begin{table}[t]
\centering
\small
\caption{Emotion recognition performances for the EEG model.}
\begin{tabular}{M{0.7cm} M{1.57cm} M{1.36cm} M{1.57cm} M{1.36cm}}
\toprule[1pt]
 \textbf{Pre-train} & \textbf{Arousal Acc.} & \textbf{Arousal F1} & \textbf{Valence Acc.} & \textbf{Valence F1}\\
\midrule[1pt]
No & 0.76$\pm7.3\mathrm{e}^{-3}$ & 0.75$\pm8.3\mathrm{e}^{-3}$ & 0.7$\pm6.4\mathrm{e}^{-3}$ & 0.7$\pm6.8\mathrm{e}^{-3}$ \\
Yes & \textbf{0.81$\pm10.7\mathrm{e}^{-3}$} & \textbf{0.80$\pm9.4\mathrm{e}^{-3}$} & \textbf{0.77$\pm9.3\mathrm{e}^{-3}$} & \textbf{0.77$\pm9.1\mathrm{e}^{-3}$} \\
\bottomrule[1pt]
\end{tabular}
\label{table:eeg_pre_nopre}
\end{table}

We first report in Table \ref{table:eeg_pre_nopre} the performances of our single-modality emotion recognizer on EEG signals, using the pre-training strategy described in section \ref{section:pretrain}, and without using pre-training. We see that the pre-training strategy improves emotion recognition performances for all metrics, for both arousal and valence. For example, the F1 score for valence goes from 0.7$\pm6.8\mathrm{e}^{-3}$ when not using pre-training, to 0.77$\pm9.1\mathrm{e}^{-3}$ when pre-training is used. This confirms that the pre-training strategy we employ is useful when processing EEG signals, and helps the Transformer learn better representations that in turn produce better results when predicting emotion. The comparisons of all metrics shown in Table \ref{table:eeg_pre_nopre} have a two-tailed P value less than $1\mathrm{e}^{-5}$, thus the difference between them can be considered to be statistically significant.

\subsection{Fused Model Results}

\begin{table}[t]
\centering
\small
\caption{Emotion recognition performances of single-modality models and of the fused model.}
\begin{tabular}{p{0.8cm} M{1.57cm} M{1.36cm} M{1.57cm} M{1.36cm}}
\toprule[1pt]
& \textbf{Arousal Acc.} & \textbf{Arousal F1} & \textbf{Valence Acc.} & \textbf{Valence F1}\\
\midrule[1pt]
\multirow{1}{1.5cm}{ECG \cite{vazquez-rodriguezTransformerBasedSelfSupervisedLearning2022}} & 0.88$\pm5.4\mathrm{e}^{-3}$ & 0.87$\pm5.4\mathrm{e}^{-3}$ & 0.83$\pm7.8\mathrm{e}^{-3}$ & 0.83$\pm7.4\mathrm{e}^{-3}$ \\
EEG & 0.81$\pm10.7\mathrm{e}^{-3}$ & 0.80$\pm9.4\mathrm{e}^{-3}$ & 0.77$\pm9.3\mathrm{e}^{-3}$ & 0.77$\pm9.1\mathrm{e}^{-3}$ \\
Fused & \textbf{0.89$\pm5.0\mathrm{e}^{-3}$} & \textbf{0.89$\pm5.0\mathrm{e}^{-3}$} & \textbf{0.85$\pm3.8\mathrm{e}^{-3}$} & \textbf{0.85$\pm3.9\mathrm{e}^{-3}$} \\
\bottomrule[1pt]
\end{tabular}
\label{table:ecg_eeg_fused}
\end{table}

In Table \ref{table:ecg_eeg_fused} we present the performances of our fused model, along with the performances of the single-modality models. Pre-training was used in the single-modality models, and the fused model employs those single-modality models as part of the fusion strategy, as described in Section \ref{section:fusion_estrategy}. We see that our late-fusion approach improves performance over single-modality models. For example, we obtain a valence accuracy in the fused model of 0.85$\pm3.8\mathrm{e}^{-3}$ compared to 0.83$\pm7.8\mathrm{e}^{-3}$ and 0.77$\pm9.3\mathrm{e}^{-3}$ when using only EEC and EEG signals respectively. Moreover, when comparing the results of the single-modality models with the fused strategy, the two-tailed P values are less than $1\mathrm{e}^{-3}$, thus the difference is statistically significant.

\begin{figure}[t]
  \centering
  \begin{tikzpicture}[scale=0.8]

\begin{axis}[
    width=0.6\textwidth,
    height=5cm,
    xlabel={(a)},
    xtick pos=left,
    ytick pos=left,
    ticklabel style = {font=\small},
    ]
    
    \addplot +[mark=none, ultra thin] table[col sep=comma] {data/Ex01_ECG_wrong.csv};
\end{axis}

\end{tikzpicture}

\begin{tikzpicture}[scale=0.8]

\begin{groupplot}[
    group style={
        group size=1 by 2,
        vertical sep=0pt,
        xlabels at=edge bottom,
        xticklabels at=edge bottom,
        },
    width=0.6\textwidth,
    height=4cm,
    xlabel={(b)},
    xtick pos=left,
    ytick pos=left,
    ticklabel style = {font=\small},
    ]
\nextgroupplot
    \addplot +[mark=none, ultra thin] table[col sep=comma] {data/Ex01_EEG_right.csv};
\nextgroupplot
    \addplot +[mark=none, ultra thin] table[col sep=comma, y index = {2}] {data/Ex01_EEG_right.csv};
\end{groupplot}

\end{tikzpicture}
  \vspace*{-7.5mm}
  \caption{Sample correctly classified by the Fused Model and the EEG model, but incorrectly classified by the ECG model. Figure (a) shows the ECG signal, and Figure (b) shows channels F7 and F3 of the EEG signal.}
  \label{fig:ecg_wrong}
\end{figure}
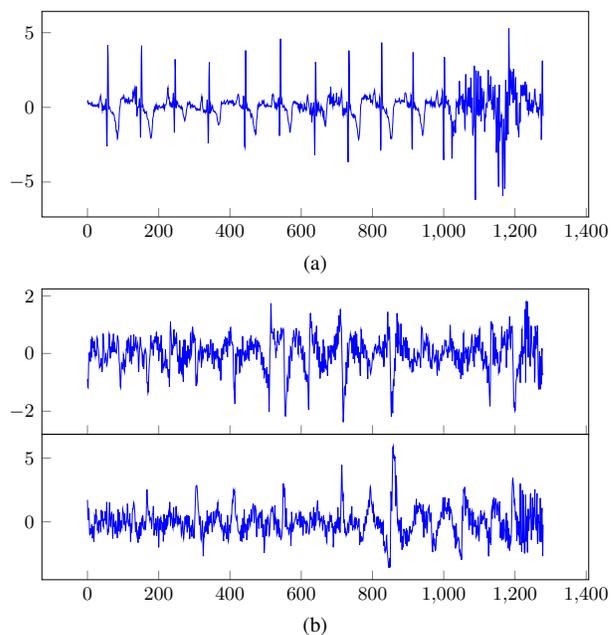

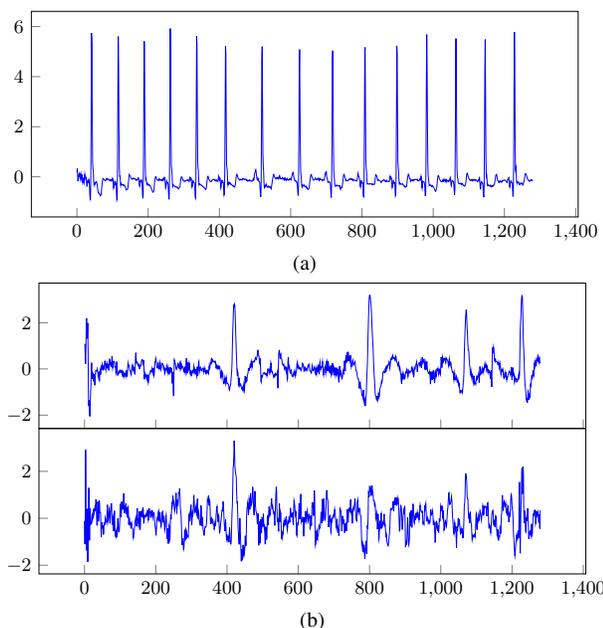
\begin{figure}[t]
  \centering
  \begin{tikzpicture}[scale=0.8]

\begin{axis}[
    width=0.6\textwidth,
    height=5cm,
    xlabel={(a)},
    xtick pos=left,
    ytick pos=left,
    ticklabel style = {font=\small},
    ]
    
    \addplot +[mark=none, ultra thin] table[col sep=comma] {data/Ex02_ECG_right.csv};
\end{axis}

\end{tikzpicture}

\begin{tikzpicture}[scale=0.8]

\begin{groupplot}[
    group style={
        group size=1 by 2,
        vertical sep=0pt,
        xlabels at=edge bottom,
        xticklabels at=edge bottom,
        },
    width=0.6\textwidth,
    height=4cm,
    xlabel={(b)},
    xtick pos=left,
    ytick pos=left,
    ticklabel style = {font=\small},
    ]
\nextgroupplot
    \addplot +[mark=none, ultra thin] table[col sep=comma] {data/Ex02_EEG_wrong.csv};
\nextgroupplot
    \addplot +[mark=none, ultra thin] table[col sep=comma, y index = {2}] {data/Ex02_EEG_wrong.csv};
\end{groupplot}

\end{tikzpicture}
   \vspace*{-7.5mm}
  \caption{Sample correctly classified by the Fused Model and the ECG model, but incorrectly classified by the EEG model. The ECG signal is shown in Figure (a) and channels F7 and F3 of the EEG signal are shown in Figure (b).}
  \label{fig:ecg_right}
\end{figure}

Figures \ref{fig:ecg_wrong} and \ref{fig:ecg_right} show two different samples used in our experiments. Figure \ref{fig:ecg_wrong}(a) and Figure \ref{fig:ecg_right}(a) show the ECG signals, and Figure \ref{fig:ecg_wrong}(b) and \ref{fig:ecg_right}(b) show channels F7 and F3 of the EEG signal. For the sample of Figure \ref{fig:ecg_wrong}, the ECG model predicts the wrong class, while the EEG model predicts the correct class. We can see that the fused model is helpful when the ECG modality makes a wrong prediction, relying on the information from the EEG signal to do the classification correctly. In Figure \ref{fig:ecg_right}, we present the same channels as in the previous example, but for a sample that is classified correctly by the ECG model and misclassified by the EEG model. Now the fused model is capable of relying on the information from the ECG signal to classify correctly this sample. Therefore, our fusion model is capable of paying attention to the right modality when one is informative and the other leads to incorrect predictions. Looking at the signals for these 2 samples, it is not obvious why misclassifications occur for one modality or the other, compared to other signals. This showcases that our model is capable of extracting meaningful hidden features in both modalities.

\subsection{Effectiveness of Pre-training in the Fused Model} \label{section:effectivness_pre}

\begin{table}[t]
\centering
\small
\caption{Fused Model: Pre-Training vs No Pre-Training}
\begin{tabular}{M{0.7cm} M{1.57cm} M{1.36cm} M{1.57cm} M{1.36cm}}
\toprule[1pt]
 \textbf{Pre-train} & \textbf{Arousal Acc.} & \textbf{Arousal F1} & \textbf{Valence Acc.} & \textbf{Valence F1}\\
\midrule[1pt]
No & 0.86$\pm4.9\mathrm{e}^{-3}$ & 0.85$\pm5.1\mathrm{e}^{-3}$ & 0.82$\pm6.5\mathrm{e}^{-3}$ & 0.81$\pm6.8\mathrm{e}^{-3}$ \\
Yes & \textbf{0.89$\pm5.0\mathrm{e}^{-3}$} & \textbf{0.89$\pm5.0\mathrm{e}^{-3}$} & \textbf{0.85$\pm3.8\mathrm{e}^{-3}$} & \textbf{0.85$\pm3.9\mathrm{e}^{-3}$} \\
\bottomrule[1pt]
\end{tabular}
\label{table:eeg_fused_pre_nopre}
\end{table}

In Table \ref{table:eeg_fused_pre_nopre}, we compare the performances of our fused model, depending on whether it uses pre-trained single-modality models, or single-modality models with no pre-training. We see that the pre-trained fused model achieves superior performance compared to the fused model with no pre-training. For example, the arousal F1 score improves from 0.85$\pm5.1\mathrm{e}^{-3}$ to 0.89$\pm5.0\mathrm{e}^{-3}$. The results shown in Table \ref{table:ecg_eeg_fused} have two-tailed P values less than $1\mathrm{e}^{-4}$, thus the difference between them is extremely statistically significant. These results indicate that the benefits obtained from pre-training single-modality models are carried over when combining them with our late-fusion strategy.

\subsection{Comparison with some baselines on AMIGOS dataset}

\begin{table}[t]
\centering
\small
\caption{Comparison with the AMIGOS Dataset Baseline}
\begin{tabular}{p{3.6cm} p{1.5cm} p{1.5cm}}
\toprule[1pt]
& \textbf{Arousal F1} & \textbf{Valence F1}\\
\midrule[1pt]
EEG \cite{correaAMIGOSDatasetAffect2018} & 0.577 & 0.564 \\
EEG (ours) & 0.80$\pm9.4\mathrm{e}^{-3}$ & 0.77$\pm9.1\mathrm{e}^{-3}$ \\
\midrule
Fused: EEG+ECG+GRS \cite{correaAMIGOSDatasetAffect2018} & 0.564 & 0.560 \\
Fused: EEG+ECG (ours) & \textbf{0.89$\pm5.0\mathrm{e}^{-3}$} & \textbf{0.85$\pm3.9\mathrm{e}^{-3}$} \\
\bottomrule[1pt]
\end{tabular}
\label{table:amigos_comp}
\end{table}

We report in Table \ref{table:amigos_comp} the performance of our models and the results reported by the authors of the AMIGOS dataset \cite{correaAMIGOSDatasetAffect2018}, that we consider as a baseline. We should note that the experimental protocol used in the baseline is different than our protocol, for example the length of their segments is 20s, thus their and our results are not directly comparable. In any case, we present those results to showcase results obtained by other works, and to see if our approach has acceptable performance. We see that our performances are much higher than the baseline, both using only EEG and also in the fused approach. The fused approach in the baseline uses Galvanic Skin Response (GSR) in addition to EEG and ECG. Our approach is performing at much higher F1 scores with one less modality, which further validates that our approach is promising.

\section{Conclusion and Perspectives}
In this work, we presented a new Transformer-based architecture with pre-training for emotion recognition on multimodal physiological signals. We experimentally showed, using the AMIGOS dataset, that our approach can predict valence and arousal with significant accuracy. In addition, we demonstrated that our late-fusion multimodal approach improves performances over single-modality. Finally, we compared the benefits of our pre-training strategy for multimodal situations. Overall, our architecture is capable of reaching state-of-the-art performance for emotion recognition.

Future works include investigating new ways to perform pre-training in multimodal situations: instead of pre-training individual modalities in independent Transformers, we can expect that using a single multimodal Transformer could lead to better performances. Indeed, this way, the model could start to incorporate information from different modalities as early as in the pre-training phase. In general, new ways of combining different modalities in a pre-trainable Transformer architecture, be it early-fusion, middle-fusion, or late-fusion, are valuable avenues of research to improve emotion recognition performances of such models. Moreover, the combination of physiological signals with more traditional modalities such as images and audio may help to better understand how pre-training and Transformer-based models behave for multimodal emotion recognition.

\vspace{0.25cm}
{\small
\textbf{Acknowledgements:} This work has been partially supported by the MIAI Multidisciplinary AI Institute at the Univ. Grenoble Alpes:  (MIAI@Grenoble Alpes - ANR-19-P3IA-0003).}

\bibliographystyle{IEEEbib}
\bibliography{refs.bib}

\end{document}